\documentclass[a4paper,fleqn]{cas-dc}

\ifdefined\LaTeXML
  \newcolumntype{L}{l}%
  \newcolumntype{R}{r}%
  \newcolumntype{C}{c}%
  \providecommand{\tblwidth}{\textwidth}%
  \providecommand{\cormark}[1]{}%
  \providecommand{\cortext}[2][]{}%
  \providecommand{\tnotemark}[1]{}%
  \providecommand{\tnotetext}[2][]{}%
  \providecommand{\credit}[1]{}%
  \providecommand{\sep}{, }%
  \providecommand{\printcredits}{}%
  \providecommand{\nonumnote}[1]{#1}%
\fi

\ifdefined\LaTeXML\else
  \ExplSyntaxOn
  \cs_set:Npn \__first_footerline:
  {
    \group_begin:
    \small
    \sffamily
    \ifnum\theblind>0\relax
    \else
      \__short_authors: :~
    \fi
    { \rmfamily \itshape Author~ accepted~ manuscript~ \textemdash{}~ Environmental~ Modelling~ \&~ Software }
    \group_end:
  }
  \ExplSyntaxOff
\fi

\usepackage[numbers]{natbib}
\usepackage[linesnumbered,lined,boxed]{algorithm2e}
\usepackage{algpseudocode}
\usepackage{multirow}
\usepackage{bbding}
\usepackage{hyphenat}
\usepackage{mathtools}
\usepackage{textcomp}
\usepackage[colorlinks]{hyperref}
\usepackage[acronym,nogroupskip]{glossaries}
\ifdefined\LaTeXML\else
  \usepackage{xurl}%
\fi
\ifdefined\LaTeXML
  \providecommand{\balance}{}%
\else
  \usepackage{balance}%
\fi
\def\H{H\xspace}
\def\W{W\xspace}
\def\Vw{V_w\xspace}
\def\thetaw{\theta_w\xspace}
\def\thetas{\theta_s\xspace}
\def\pveg{p_\text{veg}\xspace}
\def\pden{p_\text{den}\xspace}
\def\Sinit{S_\text{init}\xspace}
\def\E{E\xspace}
\def\Ehat{\hat{E}\xspace}
\def\l{l\xspace}
\def\cone{c_1\xspace}
\def\ctwo{c_2\xspace}
\def\a{a\xspace}
\def\ph{p_h\xspace}
\def\pcontinue{p_\text{continue}\xspace}
\def\pw{p_w\xspace}
\def\ps{p_s\xspace}
\def\ppropagate{p_\text{propagate}\xspace}
\def\ppropagateij{p_{\text{propagate}, i, j}\xspace}
\def\pignite{p_\text{ignite}\xspace}
\def\pignitej{p_{\text{ignite}, j}\xspace}
\def\Scurrent{S_\text{current}\xspace}
\def\Snext{S_\text{next}\xspace}
\def\c{c\xspace}
\def\fp{f_p\xspace}
\def\rfirst{r_\text{first}\xspace}
\def\rbetween{r_\text{between}\xspace}
\def\rlast{r_\text{last}\xspace}
\def\t{t\xspace}
\def\it{it\xspace}
\def\y{y\xspace}
\def\yhat{\hat{y}\xspace}
\def\A{A\xspace}
\def\Ahat{\hat{A}\xspace}

\newglossarystyle{myliststyle}{
  \setglossarystyle{list} 
  
}

\makeglossaries

\newacronym{ca}{CA}{Cellular Automata}
\newacronym{cfd}{CFD}{Computational Fluid Dynamics}
\newacronym{cuda}{CUDA}{Compute Unified Device Architecture}
\newacronym{dca}{DCA}{Differentiable Cellular Automata}
\newacronym{ga}{GA}{Genetic Algorithm}
\newacronym{gpu}{GPU}{Graphic Processing Unit}
\newacronym{iou}{IoU}{Intersection over Union}
\newacronym{ml}{ML}{Machine Learning}
\newacronym{modis}{MODIS}{Moderate Resolution Imaging Spectroradiometer}
\newacronym{mpi}{MPI}{Message Passing Interface}
\newacronym{pypi}{PyPI}{Python Package Index}
\newacronym{mse}{MSE}{Mean Squared Error}
\newacronym{bce}{BCE}{Binary Cross Entropy}

\begin{document}
\let\WriteBookmarks\relax
\def\floatpagepagefraction{1}
\def\textpagefraction{.001}

\shorttitle{PyTorchFire: A GPU-Accelerated Wildfire Simulator with Differentiable Cellular Automata}

\shortauthors{Xia et~al.}

\title[mode = title]{PyTorchFire: A GPU-Accelerated Wildfire Simulator with Differentiable Cellular Automata}

\author[1]{Zeyu Xia}[orcid=0000-0003-0234-5857]


\credit{Conceptualization, Methodology, Software, Validation, Formal analysis, Investigation, Resources, Data Curation, Writing - Original Draft, Visualization, Project administration}

\affiliation[1]{organization={University of Virginia},
    city={Charlottesville},
    postcode={VA 22904},
    country={USA}}

\author[2]{Sibo Cheng}[orcid=0000-0002-8707-2589]

\cormark[1]

\ead{sibo.cheng@enpc.fr}

\credit{Conceptualization, Methodology, Resources, Writing - Review \& Editing, Supervision, Funding acquisition}

\affiliation[2]{organization={CEREA, ENPC, EDF R\&D, Institut Polytechnique de Paris},
    state={\^Ile-de-France},
    country={France}}

\cortext[1]{Corresponding author}

\begin{abstract}
    Accurate and rapid prediction of wildfire trends is crucial for effective management and mitigation. However, the stochastic nature of fire propagation poses significant challenges in developing reliable simulators. In this paper, we introduce \texttt{PyTorchFire}, an open-access, \texttt{PyTorch}-based software that leverages \acrshort*{gpu} acceleration. With our redesigned differentiable wildfire \acrfull*{ca} model, we achieve millisecond-level computational efficiency, significantly outperforming traditional CPU-based wildfire simulators on real-world-scale fires at high resolution. Real-time parameter calibration is made possible through gradient descent on our model, aligning simulations closely with observed wildfire behavior both temporally and spatially, thereby enhancing the realism of the simulations. Our \texttt{PyTorchFire} simulator, combined with real-world environmental data, demonstrates superior generalizability compared to supervised learning surrogate models. Its ability to predict and calibrate wildfire behavior in real-time ensures accuracy, stability, and efficiency.\ \texttt{PyTorchFire} has the potential to revolutionize wildfire simulation, serving as a powerful tool for wildfire prediction and management.
\end{abstract}



\nonumnote{Author accepted manuscript. Version of record: \textit{Environmental Modelling \& Software}, vol.~188, article~106401, Apr.~2025. \href{https://doi.org/10.1016/j.envsoft.2025.106401}{doi:10.1016/\allowbreak j.envsoft.\allowbreak 2025.\allowbreak 106401}. The version of record is \textcopyright~2025 The Authors, published by Elsevier Ltd under \href{https://creativecommons.org/licenses/by/4.0/}{CC BY 4.0}.}

\maketitle

\section*{Code availability}\label{SEC:CODE}

\textbf{Name of software:} \texttt{PyTorchFire}

\textbf{Developer:} Zeyu Xia, \href{mailto:zeyu.xia@virginia.edu}{zeyu.xia@virginia.edu}

\textbf{Year first available:} 2024

\textbf{Hardware requirements:} Preferred NVIDIA GPU

\textbf{Software requirements:} Python 3.8+, preferred 3.11

\textbf{License:} MIT License

\textbf{Program language:} Python 3

\textbf{Homepage}: \url{https://github.com/xiazeyu/PyTorchFire}

\textbf{Source code:} \url{https://doi.org/10.5281/zenodo.13132218}

\textbf{PyPI package:} \url{https://pypi.org/project/pytorchfire}

\textbf{Document:} \url{https://pytorchfire.readthedocs.io}

\textbf{Dataset:} \url{https://doi.org/10.17632/nx2wsksp9k.1}

\begin{table*}[pos=h]
    \centering
    {\Large \textbf{Main Notations} \\}
    \vspace{1em}
    \begin{tabular}{rl}
        \toprule
        Notation                       & Description                                                          \\
        \midrule
        \multicolumn{2}{l}{\textit{Notations for Dataset}}                                                    \\
        \midrule
        \(\H, \W\)                     & Height and width of data                                             \\
        \(\Vw\)                        & Wind velocity                                                        \\
        \(\thetaw\)                    & Wind direction                                                       \\
        \(\thetas\)                    & Slope to neighboring cells                                           \\
        \(\pveg\)                      & Scaling factor for vegetation type                                   \\
        \(\pden\)                      & Scaling factor for vegetation density                                \\
        \(\Sinit\)                     & Initial fire state                                                   \\
        \(\E, \Ehat\)                  & Altitude of a cell and its neighbors                                 \\
        \(\l\)                         & Side length (resolution) of dataset                                  \\
        \midrule
        \multicolumn{2}{l}{\textit{Notations for Wildfire Cellular Automata}}                                 \\
        \midrule
        \(\cone\)                      & Scaling factor for wind velocity                                     \\
        \(\ctwo\)                      & Scaling factor for wind direction                                    \\
        \(\a\)                         & Scaling factor for ground elevation                                  \\
        \(\ph\)                        & Base burning probability                                             \\
        \(\pcontinue\)                 & Probability of continued burning                                     \\
        \(\pw\)                        & Scaling factor for wind                                              \\
        \(\ps\)                        & Scaling factor for slope                                             \\
        \(\c\)                         & Constant for normalization function                                  \\
        \(\fp\)                        & Probability-like normalization function                              \\
        \(\ppropagateij, \ppropagate\) & Probability for cell \(i\) to propagate fire to cell \(j\)           \\
        \(\pignitej, \pignite\)        & Probability of cell \(j\) becoming ignited                           \\
        \(\Scurrent\)                  & Current fire state                                                   \\
        \(\Snext\)                     & Next fire state                                                      \\
        \midrule
        \multicolumn{2}{l}{\textit{Notations for Parameter Calibration Trainer}}                              \\
        \midrule
        \(\rfirst, \rbetween, \rlast\) & Constants for step rings attached to accumulator                     \\
        \(\t\)                         & Time step                                                            \\
        \(\it\)                        & Iteration step                                                       \\
        \(\y, \yhat\)                  & Observed fire region binary label and predicted ignition probability \\
        \(\A, \Ahat\)                  & Observed and predicted fire region                                   \\
        \bottomrule
    \end{tabular}
\end{table*}

\section{Introduction}\label{SEC:INTRO}

There has been an increasing number of devastating effects caused by human-induced climate change. One of its outcomes is the more frequent occurrence of wildfires, which are larger in both scale and duration worldwide~\cite{xuWildfiresGlobalClimate2020}. There are approximately 60,000 undesirable wildfire events (not intentionally set by humans for forest health and safety of nearby communities) with an average burned area of 7.02 million acres annually in the last decade just in the United States~\cite{WildfiresAcresNational}.  These fire events have not only put properties at high risk of being damaged by wildfires~\cite{auerIncomeInsurabilityFactors2022}, but also pose a significant threat to human health~\cite{chenMortalityRiskAttributable2021}.

Having an accurate and effective prediction, especially in the early stages of a fire event, is crucial for decision making in firefighting and evacuation strategies, as well as for short-term emergency response and long-term fire risk assessment~\cite{hansonPotentialPromisePhysicsbased2000}. Numerous studies have focused on predicting fire spread using various models, such as the semi-empirical model~\cite{finneyFlamMapFireMapping2023,doyleSimFire2024}, \gls{cfd}-based models~\cite{valeroMultifidelityPredictionWildfire2021}, data-driven models~\cite{maffeiPredictingForestFires2019}, and \gls{ca}-based models~\cite{hernandezencinasSimulationForestFire2007,alexandridisCellularAutomataModel2008,papadopoulosComparativeReviewWildfire2011,trucchiaPROPAGATOROperationalCellularAutomata2020,lopez-de-castroFireSpottingModellingOperational2024}. However, these models typically utilize the CPU only, which is not fast enough to provide real-time predictions and cannot make predictions in a reasonable time when the fire scale is large or the resolution is high. Recent advances in neural networks~\cite{XiandaOpenStereoAComprehensive2024,xiaAccurateIdentificationMeasurement2023,wenmingSingingVoiceDetection2023,zhongInvolutionBasedSpeech2021} have introduced a novel approach to the construction of wildfire prediction models. Numerous studies~\cite{chengParameterFlexibleWildfire2022,chengDataDrivenSurrogateModel2022,jainReviewMachineLearning2020} have leveraged \gls{ml} surrogate modeling to enhance the efficiency of wildfire simulations. Nevertheless, these methods exhibit significant limitations regarding generalizability. They are typically trained in specific ecoregions, rendering them applicable solely to those particular regions and unsuitable for deployment in diverse ecoregions with varying geophysical conditions. There has been recent research leveraging \gls{mpi} for parallelization on CPUs~\cite{liXCLiParallel_CellularAutomaton_Wildfire2018} and utilizing \gls{gpu} acceleration~\cite{hoangWildfireSimulationGPU2008,ntinasParallelFuzzyCellular2017,denhamUsingEfficientParallelization2018} to enhance wildfire prediction. However, these implementations are frequently developed in C, \gls{cuda}, or MATLAB programming language, posing challenges for integration into existing wildfire prediction systems due to specific system requirements  or the necessity for commercial licenses. It may also cause trouble for the user to use user-provided datasets. In contrast, our proposed approach aims to achieve portability such that, by merely altering the device string, the same software can be executed across different architectures, including CPUs and \glspl{gpu} from various vendors.

Another significant challenge in achieving reliable wildfire prediction is parameter calibration or model correction. As summarized in~\cite{papadopoulosComparativeReviewWildfire2011}, a variety of algorithms~\cite{alexandridisCellularAutomataModel2008, chengDataDrivenSurrogateModel2022, yassemiDesignImplementationIntegrated2008} have been designed to forecast the progress of regional fires based on local characteristics such as terrain, wind and fuel. However, due to the complexity and chaotic nature of wildfire systems, prediction models often lack accuracy, and the results heavily depend on the initial parameters~\cite{alessandriParameterEstimationFire2021}. Parameter calibration is crucial for wildfire models to reduce prediction bias. Despite traditional brute-force methods~\cite{alexandridisCellularAutomataModel2008, denhamUsingEfficientParallelization2018}, several studies focus on parameter identification using \gls{ga}~\cite{denhamUsingEfficientParallelization2018,ntinasParallelFuzzyCellular2017}, variational data assimilation~\cite{chengParameterFlexibleWildfire2022}, or the ensemble Kalman filter~\cite{zhangStateparameterEstimationApproach2019}. However, these methods are either mathematically unstable, computationally expensive or require a large amount of data, making it challenging to perform parameter calibration on a large scale or in real-time. It is also worth mentioning that, due to the stochastic nature of wildfire modellings, different runs yield varying outcomes, complicating parameter calibration.

In this paper, we developed \texttt{PyTorchFire}, a \texttt{PyTorch}-based simulator designed to provide a fast and reliable wildfire simulation platform capable of self-correction using real-world observation data of wildfires. Our simulator incorporates a redesigned, \gls{dca} model that is efficient and supports real-time parameter calibration with excellent generalizability. With the support of \texttt{PyTorch}, we achieve high computing efficiency without compromising scalability and portability.\ \gls{gpu} acceleration not only allows us to perform single simulation steps at the millisecond level but also enables large-scale simulations and higher resolutions that were previously challenging. The probability of arrival can be computed using an ensemble of numerous \gls{ca} runs. However, tracking gradients from these ensembles is challenging, and parameter calibration can be computationally intensive. By averaging these runs, we can effectively leverage the model's stochastic nature. Our model can also conduct real-time parameter calibration using gradient descent, allowing it to self-correct with the latest real-world data.

We've first tested the prediction running time of \texttt{PyTorchFire} and \texttt{MPI-CA} (modified from~\cite{liXCLiParallel_CellularAutomaton_Wildfire2018}). It turns out \texttt{PyTorchFire} significantly outperforms \texttt{MPI-CA} and scales well on \glspl{gpu}. Then, we tested the parameter calibration performance on actual locations in California. We first calibrated fire parameters on 32 simulated fire events across three locations (Bear 2020, Brattain 2020, and Pier 2017), achieving excellent results compared to the targets. Finally, we performed parameter calibration using 2 recent real-world fire events (Bear fire in 2020 and Pier fire in 2017) in California. The results indicate that \texttt{PyTorchFire} can accurately mimic real fire events.

As our work functions similarly to traditional \gls{ml} models, we will use some terms from \gls{ml} to describe the behavior of our model. Specifically, we will use the terms \textit{forward propagation} and \textit{prediction} to describe the process of simulating the fire event. We will also use \textit{backward propagation} and \textit{training} to describe the process of calibrating the fire parameters.

Table~\ref{TBL:1} compares the capabilities of various well-known fire simulators. To the best of our knowledge, we are the first to apply \acrfull{dca} to wildfire simulation and to demonstrate its performance using real-world fire events.

\begin{table*}[width=\textwidth,cols=6,pos=h]
    \caption{Comparison of various fire simulators' capabilities.}\label{TBL:1}
    \begin{tabular*}{\tblwidth}{@{} LLLLLL@{} }
        \toprule
        & GPU-acceleration & Generalizable & Differentiable & Real-time calibration & Test on real fire event \\ \midrule
        SimFire~\cite{doyleSimFire2024} & \XSolidBrush & \Checkmark & \XSolidBrush & \XSolidBrush & \XSolidBrush \\
        Flammap~\cite{finneyFlamMapFireMapping2023} & \XSolidBrush & \Checkmark & \XSolidBrush & \XSolidBrush &  \Checkmark \\
        MPI-CA~\cite{liXCLiParallel_CellularAutomaton_Wildfire2018,alexandridisCellularAutomataModel2008,trucchiaPROPAGATOROperationalCellularAutomata2020} & \XSolidBrush & \Checkmark & \XSolidBrush & \XSolidBrush &  \Checkmark \\
        ML Models~\cite{chengDataDrivenSurrogateModel2022} & \Checkmark & \XSolidBrush & \Checkmark & \Checkmark &  \Checkmark \\
        FCA~\cite{ntinasParallelFuzzyCellular2017} & \Checkmark & \Checkmark & \XSolidBrush & \Checkmark (unstable) & \XSolidBrush \\
        CUDAFires~\cite{denhamUsingEfficientParallelization2018} & \Checkmark & \Checkmark & \XSolidBrush & \Checkmark  (unstable) & \XSolidBrush \\
        \textbf{PyTorchFire} & \Checkmark & \Checkmark & \Checkmark & \Checkmark &  \Checkmark  \\
        \bottomrule
    \end{tabular*}
\end{table*}

The structure of the rest of this paper is outlined as follows: Section~\ref{SEC:METHOD} presents a detailed description of our software design, including both wildfire \gls{dca} and algorithm for parameter calibration. The experiment design and results analysis on both simulated and real-world fire events are discussed in Section~\ref{SEC:RESULT}. The paper concludes in Section~\ref{SEC:CONCLUSION} with a synthesis of the key insights gathered and potential future work. The Appendix section provides additional information on the datasets, experiment environment, and learning rates used in this study.

\section{Methods}\label{SEC:METHOD}

\subsection{Inputs, outputs, and parameters}

Our \gls{dca} model requires several tensors as input. If a specific tensor is missing, an empty tensor will be used instead. All required inputs, including environmental data and initial fire spread, are described in the following list, with \(\H\) denoting the height from the rasterized ecoregion, and \(\W\) denoting the width:

\begin{itemize}
    \item \textbf{Wind impact}: There are two input tensors related to wind:
          \begin{itemize}
              \item \textbf{Wind velocity} \(\Vw\): \(\H\times\W\) float tensor specifying the average wind speed per cell, measured in m/s.
              \item \textbf{Wind direction} \(\thetaw\): \(\H\times\W\) float tensor indicating the average wind direction per cell, starting from East and progressing counterclockwise, measured in degrees.
          \end{itemize}
    \item \textbf{Slope} \(\thetas\): \(\H\times\W\times3\times3\) float tensor representing the slope from the current cell to its neighbors, measured in degrees.
    \item \textbf{Canopy} \(\pveg\): \(\H\times\W\) float tensor representing a probability scaling factor for the type of vegetation.
    \item \textbf{Density} \(\pden\): \(\H\times\W\) float tensor describing a probability scaling factor for the density of vegetation.
    \item \textbf{Initial fire} \(\Sinit\): \(\H\times\W\) boolean tensor specifying the initial burning state, serving as the starting condition for the simulation.
\end{itemize}

The slope \(\thetas\) in our model can be calculated from the altitude map using Equation~\ref{EQN:1}, where \(\E\) and \(\Ehat\) are the altitudes of current cell and its neighbors, respectively, and \(\l\) is the side length of the cell, often equal to the resolution of the dataset.

\begin{equation} \label{EQN:1}
    \thetas = \arctan\left(\frac{\E - \Ehat}{k \cdot \l}\right)
\end{equation}

Here,

\[
    k = \begin{cases}
        1        & \text{for adjacent cells} \\
        \sqrt{2} & \text{for diagonal cells}
    \end{cases}
\]

The output is a 2-channel boolean tensor of shape \(2 \times \H \times \W\), representing the predicted burned area. The first channel indicates if a cell is burning, while the second channel indicates if a cell is burnt out. If the user is performing parameter calibration, the calibrated parameters will be included as additional output. Users can access an internal function to obtain a real-time ignition map \(\pignite\), where each element represents the probability that the corresponding cell is ignited.

In our model, there are five parameters that control the fire spreading process. All of them are floating-point numbers represented as tensors. During forward propagation, these parameters are treated as inputs only. During backward propagation, they are treated as initial values and are updated accordingly, except for \(\pcontinue\). After all epochs, the optimal values with the best metrics are output as the final calibrated parameters.

The parameters we used are as follows:

\begin{itemize}
    \item \(\cone\): The scaling factor for wind velocity.
    \item \(\ctwo\): The scaling factor for wind direction.
    \item \(\a\): The scaling factor for slope.
    \item \(\ph\): The base probability that a fire propagates from a burning cell to an adjacent cell under normal conditions.
    \item \(\pcontinue\): The probability that a burning cell continues to burn in the next time step.
\end{itemize}

The \(\pcontinue\) parameter was introduced by Li's work~\cite{liXCLiParallel_CellularAutomaton_Wildfire2018}. We adopted this parameter to enable adjustment to the shape of the fire boundary. Users can disable this feature by setting \(\pcontinue=0\).

Depending on the dataset type, end-users may need to perform preprocessing to ensure compatibility with our code. Utilities for common conversions are provided with our code. Our program is designed to be modular and flexible with map reoslutions, enabling users to easily build subclasses on top of it. This modularity allows users to run simulations on their specific datasets or implement revised algorithms with ease.

\subsection{Differentiable wildfire cellular automata}\label{SEC:CA}

\gls{ca}, first popularized through Conway's Game of Life~\cite{bakSelforganizedCriticalityGame1989}, have proven to be effective tools for simulating complex stochastic processes using simple transition rules. In \texttt{PyTorchFire}, we build on the work of Alexandridis et~al.~\cite{alexandridisCellularAutomataModel2008} to create a \gls{dca} model for wildfire spread prediction.

\begin{figure}[pos=htbp]
    \centering
    \includegraphics[width=\columnwidth]{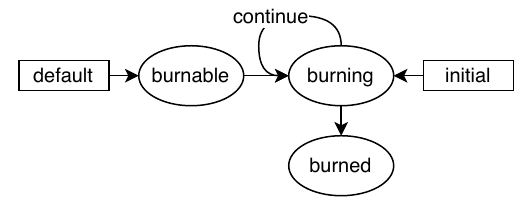}
    \caption{State transition diagram illustrating dynamic changes between wildfire states.}
    \label{FIG:1}
\end{figure}

\textbf{State and transition rules.} The model is based on a 2D grid of cells, each representing a square unit of geological area with one of three states: burnable, burning, or burned. As shown in Figure~\ref{FIG:1}, with the model being repeatedly updated in discrete time steps, each cell's state \(\Scurrent\) at time step \(\t\) will be updated to state \(\Snext\) depending on the states of its neighbors at the previous time step \(\t-1\). The transition rules are as follows:

\begin{itemize}
    \item A cell will have a default state of burnable.
    \item Initial ignition cells are set to burning.
    \item A burnable cell will ignite with probability \(\pignite\) if at least one of its neighbors is burning.
    \item A burning cell will remain burning with probability \(\pcontinue\); otherwise, it will burn out and become burned.
    \item A burned cell will always remain burned.
\end{itemize}

This state and transition design remains clear yet effective and can be easily implemented using a tensor with two boolean channels: burning and burned, which requires only 2 bits of space to store per cell. The original work~\cite{alexandridisCellularAutomataModel2008} include a state of `no forest fuel', which is considered redundant in our proposed model. This state can be represented by assigning the corresponding area a very small \(\pveg\), resulting in a fire propagation probability of zero.

\textbf{Ignition probability calculation.} To make the \gls{ca} model differentiable, we need to replace the stochastic ignition probability calculation with a differentiable function~\cite{martinDifferentiableCellularAutomata2017,mordvintsevGrowingNeuralCellular2020}.

The original \gls{ca} model~\cite{alexandridisCellularAutomataModel2008} provides a probability \(\ppropagate\) denoting a cell's likelihood to propagate fire to its neighbors, calculated as a product of various scaling factors, given by:

\begin{equation} \label{EQN:2}
    \ppropagate = \ph (1 + \pveg) (1 + \pden) \pw \ps
\end{equation}

Here, \(\ph\) is the base probability of a cell igniting, \(\pveg\) is the vegetation factor, \(\pden\) is the density factor, \(\pw\) is the wind factor, and \(\ps\) is the slope factor.

Specifically, \(\pw\) can be calculated by:

\begin{equation} \label{EQN:3}
    \pw = \exp(\cone \Vw) \exp(\ctwo \Vw (\cos(\thetaw) - 1))
\end{equation}

where \(\thetaw\) is the wind direction towards its neighbors, and \(\Vw\) is the wind velocity. The wind velocity constant \(\cone\) and wind direction constant \(\ctwo\) are derived from experimental data.

The scaling factor for slope \(\ps\) can be calculated by:

\begin{equation} \label{EQN:4}
    \ps = \exp(\a \thetas)
\end{equation}

where \(\a\) is a constant for the slope factor derived from experimental data, and \(\thetas\) is the slope angle in degrees.

To apply non-local processes such as spotting fire~\cite{lopez-de-castroFireSpottingModellingOperational2024}, we propose performing in-place operations after each simulation step. Although these fire points ignited by the spotting effect cannot be directly tracked on the computation map, they can be treated as initial ignition points. The remaining pipeline can then seamlessly execute fire prediction and parameter calibration without any modifications.

As the product of scaling factors in Eq.~\ref{EQN:2} may exceed 1, causing issues when applying transformations as probabilities, we propose a probability-like normalization function \(\fp\) to normalize the propagation probability \(\ppropagate\), defined as:

\begin{equation} \label{EQN:5}
    \fp(x) = \tanh(\c \cdot x)
\end{equation}

where \(\c\) is a hyperparameter.

The function is the optimal one selected from several candidates, where \( c \) is a parameter that influences the behavior of the functions:

\begin{align}
    f(x) & = -c^{-x} + 1, \label{EQN:6}       \\
    f(x) & = -{(x + 1)}^{-c} + 1, \label{EQN:7} \\
    f(x) & = \tanh(c x). \label{EQN:8}
\end{align}

We aim to find the best differentiable function that preserves most valid values and can adjust invalid values to the valid range \([0,1]\). Given that \(\ppropagate\) produces non-negative output, we focus on correcting the positive invalid input. Using the Nelder-Mead algorithm, we fitted these functions within the range \([0.2, 0.8]\) to determine the optimal value of \( c \) for each function, and compared the sum of squared differences in this interval. Ultimately, we selected the optimal function with the minimal sum of squared differences from Equation~\ref{EQN:8} where \(\c = 1.1486328125\).

\begin{figure}[pos=htbp]
    \centering
    \includegraphics[width=\columnwidth]{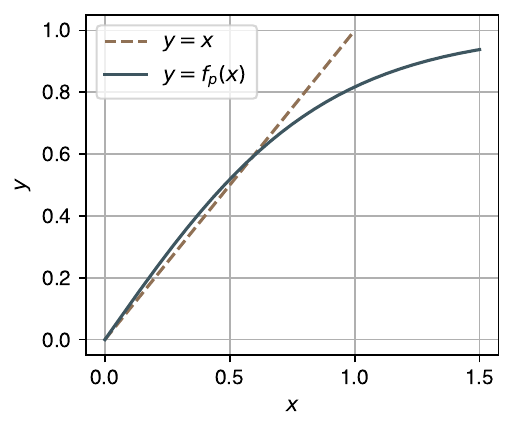}
    \caption{Comparison of our probability-like normalization function \(y=\fp(x)\) with \(y=x\), illustrating its effectiveness in maintaining valid probabilities within \([0, 1]\) and smoothly adjusting values outside this range.}
    \label{FIG:2}
\end{figure}

As shown in Figure~\ref{FIG:2}, our function is nearly linear in the range of \([0, 0.7]\), and as \(x\) approaches positive infinity, \(\fp\) approaches 1, which perfectly meets our expectations.

\begin{figure}[pos=htbp]
    \centering
    \includegraphics[width=\columnwidth]{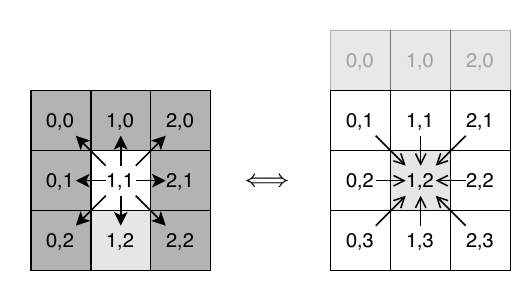}
    \caption{Propagation of \(\ppropagate\) to neighboring cells in a Moore neighborhood and the application of the inclusion-exclusion principle to determine \(\pignite\).}
    \label{FIG:3}
\end{figure}

Since Eq.~\ref{EQN:2} only provides the propagation probabilities \(\ppropagate\), it is challenging to apply gradient descent directly. Introducing a per-cell ignition probability \(\pignite\) is highly beneficial, as it can be calculated by aggregating \(\ppropagate\) from a cell's Moore neighbors. As illustrated in Figure~\ref{FIG:3}, given that the fire model assumes all ignition events from the eight neighbors are independent and identically distributed, and our normalized propagation probabilities \(\fp(\ppropagate)\) now represent true probabilities, we can apply the inclusion-exclusion principle to determine the probability of a cell being ignited by its neighboring cells. The formula is given by:

\begin{equation} \label{EQN:9}
    \pignitej = 1 - \prod_{i=1}^{8} (1 - \ppropagateij)
\end{equation}

where \(\pignitej\) is the probability of cell \(j\) being ignited, \(i\) represents cells adjacent to the current cell, and \(\ppropagateij\) is the probability of cell \(j\) being ignited by its neighbor \(i\).

\textbf{Data flow}

\begin{figure}[pos=htbp]
    \centering
    \includegraphics[width=\columnwidth]{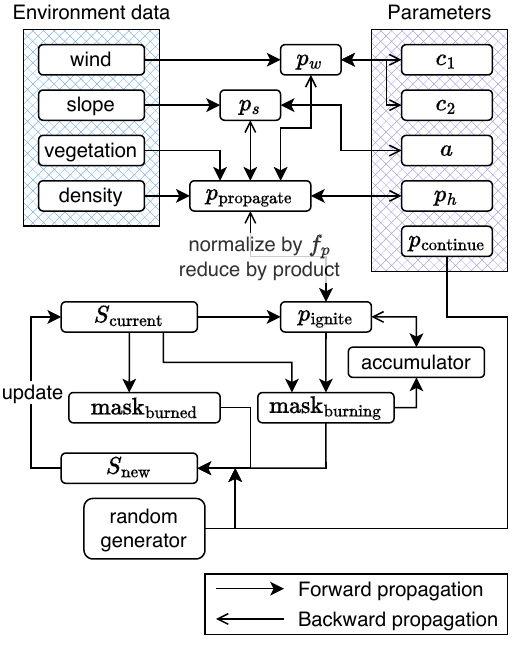}
    \caption{Flow chart depicting data flow and dependencies in \texttt{PyTorchFire}, emphasizing connections between data, parameters, and outputs.}
    \label{FIG:4}
\end{figure}

Figure~\ref{FIG:4} illustrates the data flow for a time step computation of our \gls{dca} in \texttt{PyTorchFire}. By following this data flow, the model state is updated from the current state \(\Scurrent\) to a new state \(\Snext\). The data flow direction varies depending on whether it is performing a prediction task or a parameter calibration task. Environmental data and parameters are used to calculate \(\ppropagate\) for each cell. This value is processed through our probability-like normalization function \(\fp\) to ensure it is a valid probability for subsequent steps. The \(\fp(\ppropagate)\) values of burning cells are then selected for a reduction operation to determine the burning probability \(\pignite\) for each cell. These probabilities are compared against two random matrices drawn from a uniform distribution, and based on the comparison results, the cell states are updated from \(\Scurrent\) to \(\Snext\). Finally, the \(\pignite\) values of newly ignited cells are accumulated for use in the back-propagation process during the parameter calibration task. Details about parameter calibration will be discussed in the following sections.

\subsection{Parameter calibration on the fly}

Parameter calibration based on real-time satellite observations is often necessary for wildlife predictive models to adjust parameters for more accurate future predictions. This process enhances the model's ability to predict affected areas more precisely, thereby improving its usability.

\textbf{Introduction to \gls{dca}.} The concept of performing gradient descent on \gls{dca} originates from Martin's preprint~\cite{martinDifferentiableCellularAutomata2017}. However, Martin's work focused on a single parameter and was applied in a one-dimensional context. In our research, we extended this approach to a two-dimensional space and experimented with real-world wildfire scenarios. Instead of using a simplistic fire transition model, we enabled gradient computation on a more complex model that incorporates real-world landscapes, dynamic wind conditions, and other critical factors.

\textbf{Back-propagation algorithm.} Performing parameter calibration for a stochastic \gls{dca} necessitates substantial effort to stabilize the calibration process. To address this, we developed a comprehensive algorithm based on the widely-used gradient descent algorithm, as illustrated in Algorithm~\ref{ALG:1}. The algorithm incorporates three nested loops: epoch step, iteration step \(\it\), and time step \(\t\).

The outermost epoch loop facilitates multiple complete calibrations using the entire dataset. This loop ensures that the model undergoes extensive training across several iterations, enhancing the robustness of the calibration process.

Within the epoch loop, the iteration loop processes individual data points from the dataset. Given that real-world data is often obtained in a discrete manner, this loop calibrates the parameters using all available observations, thereby achieving a temporal fire similarity.

The innermost step loop is responsible for simulating the fire using the most recently calibrated parameters each time they are updated. This approach accelerates the training process by continuously refining the parameters.

To ensure stability in the training process, we employ techniques such as using the same seed for each epoch. Additionally, after each calibration, we clip the parameters to a proper range, maintaining numerical stability.

In terms of optimizer selection, we found that AdamW outperforms SGD in convergence speed and stability. Therefore, we selected AdamW as our optimizer. For learning rate selection, we tested values ranging from 1 to \(10^{-4}\). We found that learning rates larger than \(10^{-1}\) do not converge, while learning rates smaller than \(10^{-3}\) converge extremely slowly. Consequently, we chose a learning rate of \(5 \times 10^{-3}\) for the AdamW optimizer. As the optimal learning rate varies depending on the data, we recommend users to try multiple values and use the best one.

\begin{algorithm*}
    \DontPrintSemicolon
    \caption{Parameter calibration using gradient descent}\label{ALG:1}
    \KwData{wind \(\Vw, \thetaw\), slope \(\thetas\), vegetation \(\pveg, \pden\), initial fire \(\Sinit\), observed fire \texttt{targets}}
    \KwIn{maximum epochs \texttt{max\_epochs}, update interval \ \texttt{steps\_update\_interval}, initial constants \(\cone, \ctwo, \a, \ph, \pcontinue\)}
    \KwOut{calibrated model parameters \(\cone, \ctwo, \a, \ph\)}

    Initialize wildfire model \texttt{model(\(\Vw,\thetaw,\thetas,\pveg,\pden,\Sinit,\cone,\ctwo,\a,\ph, \pcontinue\))}\;
    Initialize optimizer \texttt{optimizer}()\;
    \tcp{Set max\_iterations to the number of available data points in the dataset}
    \texttt{max\_iterations} \(\longleftarrow\) data.count()\;

    \For{\( epoch \leftarrow 1\) \KwTo \texttt{max\_epochs}}{
        \tcp{Reset the model to the initial state, and get a new random seed}
        \texttt{model}.reset()\;
        \tcp{Save the seed to stabilize training in current epoch}
        \texttt{epoch\_seed} \(\longleftarrow\) \texttt{model}.seed\;

        \For{\( iteration \leftarrow 1\) \KwTo \texttt{max\_iterations}}{
            \texttt{iter\_max\_steps} \(\longleftarrow\) \textit{iteration} * \texttt{steps\_update\_interval}\;

            \For{\( step \leftarrow 1\) \KwTo \texttt{iter\_max\_steps}}{
                \tcp{Update wind if needed}
                \texttt{model}.wind \(\longleftarrow\) data.wind[iteration]\;
                \texttt{model}.compute(attach=check\_if\_attach(\textit{step}, \texttt{iter\_max\_steps}))\;
            }

            \texttt{outputs} \(\longleftarrow\) \texttt{model}.accumulator\;
            \texttt{targets} \(\longleftarrow\) data.targets[iteration]\;
            \texttt{loss} \(\longleftarrow\) criterion(\texttt{outputs}, \texttt{targets})\;
            \tcp{Compute gradients}
            \texttt{loss}.backward()\;
            \tcp{Update weights}
            \texttt{optimizer}.step()\;
            \tcp{Reset gradients}
            \texttt{optimizer}.zero\_grad()\;

            \textbf{with} torch.no\_grad()\;
            \Indp
            \tcp{Clip the parameters to avoid numerical instability}
            \texttt{model}.a.clamp\_(min=0.0, max=1.0)\;
            \texttt{model}.c\_1.clamp\_(min=0.0, max=1.0)\;
            \texttt{model}.c\_2.clamp\_(min=0.0, max=1.0)\;
            \texttt{model}.p\_h.clamp\_(min=0.2, max=1.0)\;
            \Indm
            \tcp{Compute from beginning using new parameters}
            \texttt{model}.reset(seed=\texttt{epoch\_seed})\;
        }
    }
\end{algorithm*}

\textbf{Loss function design.} Considering that our model performs parameter calibration by minimizing the difference between simulated fire spread and observed fire spread, we need to design an appropriate loss function to measure the difference between the two in terms of shape and overall scale. We propose the following loss function to quantify the difference between the simulated and observed fire spread:

\begin{equation} \label{EQN:10}
    \begin{aligned}
        L(\y, \yhat) = \frac{1}{N} \sum_{i=1}^{N} & \left( - \left[ \y_i \cdot \log(\sigma(\yhat_i)) \right. \right.        \\
                                                  & \left. + (1 - \y_i) \cdot \log(1 - \sigma(\yhat_i)) \right]             \\
                                                  & \left. + {(\text{AvgPool2d}(\y_i) - \text{AvgPool2d}(\yhat_i))}^2 \right)
    \end{aligned}
\end{equation}

where \(\sigma(x) = \frac{1}{1 + \exp(-x)}\),\(y\) denote the binary label of the observed fire region and \(\yhat\) represents the probability of predicted ignition.

The loss function is designed to be differentiable, making it suitable for the optimization process. It comprises two parts: the \gls{bce} with logits loss, which measures the difference in fire spread shapes as a binary classification task, and the \gls{mse} loss after applying a 2D average pooling, which measures the difference in fire scale as a regression task. Our trials indicate that the \gls{bce} component effectively reconstructs the fire shape, while the \gls{mse} component accurately mimics the fire scale by approximating the number of affected cells.

To accelerate \gls{bce} loss computation, we first crop the region to the union of non-zero predictions and targets, then apply the \gls{bce} with logits loss to this region. For the \gls{mse} loss, we use average pooling with a window size of 4 and a stride of 4, allowing us to compute the expectation in each region and subsequently measure the \gls{mse} from these averages. Through experiments, we found that larger window sizes result in smaller absolute \text{MSE} values. While smaller \gls{mse} values tend to improve spatial similarity, they also increase the potential for overfitting, given the stochastic nature of the simulator.

\textbf{Step rings for back-propagation.} When performing parameter calibration, gradients can only be computed from ignited cells. As shown in Equation~\ref{EQN:9}, only ignited cells have \(\ppropagate\) contributing to \(\pignite\), so the chain rule can only be used to compute gradients of parameters from these cells. This presents a challenge: tracking gradients for all ignited cells requires substantial memory, which can slow down gradient back-propagation and potentially lead to out-of-memory errors. Conversely, tracking too few cells can result in an unevenly calibrated landscape, leading to poor performance due to underestimation or overestimation.

\begin{figure*}[pos=htbp]
    \centering
    \includegraphics[width=\textwidth]{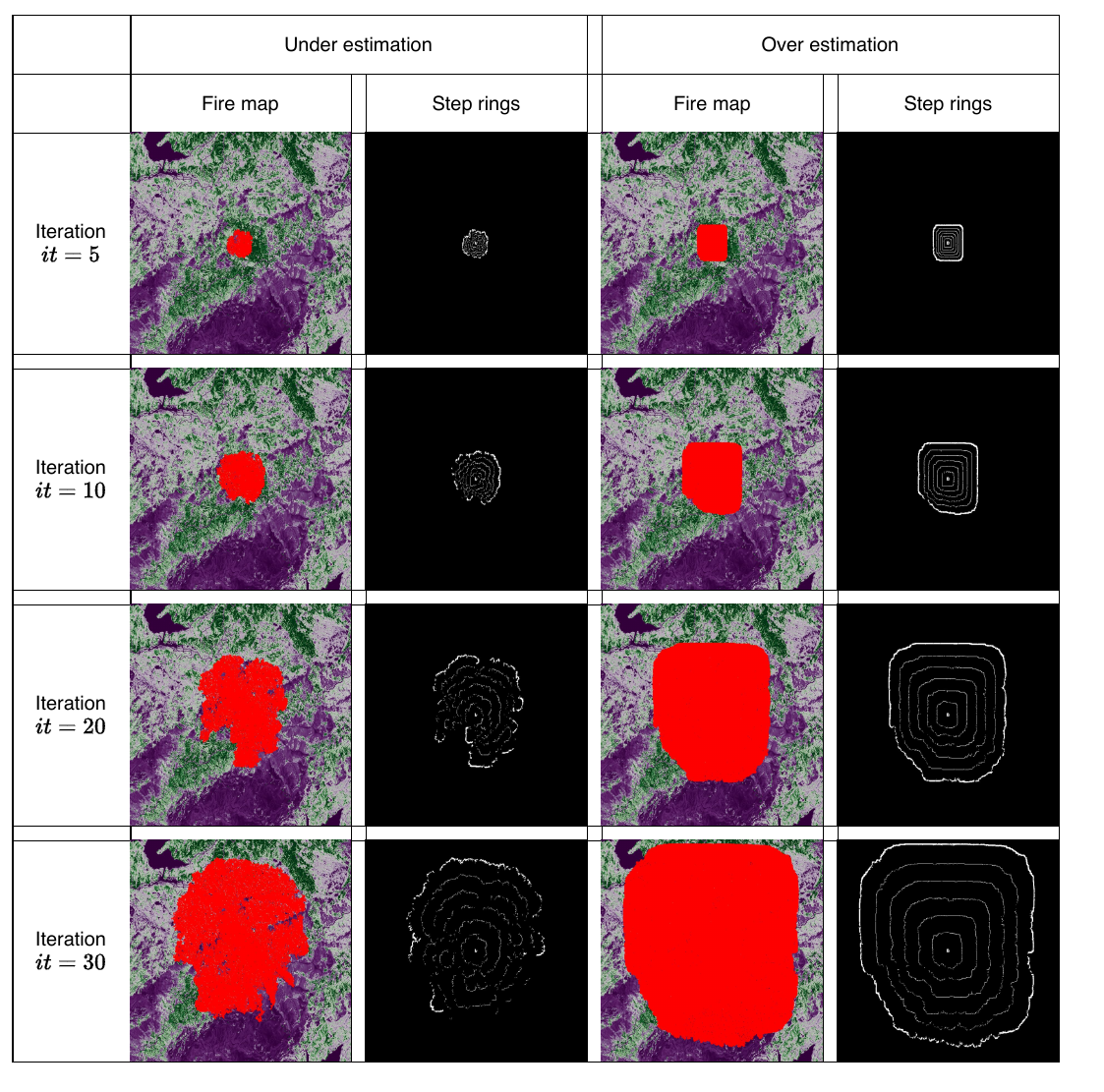}
    \caption{Selected rings of cells for back-propagation, illustrating the balance between memory usage and diversity cell coverage.}
    \label{FIG:5}
\end{figure*}

To address this issue, we proposed a novel update technique called `update by step rings'. When forming the computation graph, we always recompute the entire simulation from the beginning using the latest parameters. During this process, we determine which steps to attach to the computation graph. This is achieved by selecting the cells ignited from the first \(\rfirst\) steps, last \(\rlast\) steps, and \(\rbetween\) steps distributed evenly in between. As shown in Figure~\ref{FIG:5}, this results in a selection of multiple cells in the shape of rings, representing ignited cells from different steps. These cells are then added to an accumulator. For those cells that do not need to be connected with the computation graph, they are treated as constants and added to the accumulator as well for loss computation. This approach balances memory usage and gradient computation efficiency despite the actual state of the fire.

Through trials, we observed that increasing \(\rlast\) yields better results. Increasing \(\rbetween\) also improves performance, but not as significantly as \(\rlast\). Increasing \(\rfirst\) only has a slight impact on calibration results.

\textbf{Software description.}

To enhance accessibility, we have published our package on \gls{pypi}. Users can easily install the package by running \verb|pip install pytorchfire|. A Quick Start guide and comprehensive documentation are available online at \url{https://pytorchfire.readthedocs.io/}.

Our package relies on several essential libraries. The code is primarily based on \texttt{PyTorch}~\cite{anselPyTorchFasterMachine2024}, with version 2.0.0 or higher being preferred to ensure optimal performance. Einops~\cite{rogozhnikovEinopsClearReliable2021} is employed to facilitate readable tensor operations. In our experiments, we used h5py~\cite{colletteH5pyH5py2024} for dataset management. We recommend running our code on either \gls{cuda} or CPU platforms.

\section{Numerical Results}\label{SEC:RESULT}

In this section, as a proof of concept, we present and analyze the numerical results from \texttt{PyTorchFire} predictions, calibrated predictions using simulated data, and satellite observations.

\textbf{Experiment setup and study areas.} To evaluate the prediction performance, we assess the running time (after warm-up) of 300 time steps predictions on a simulated geographical environment. For parameter calibration performance, we selected two metrics: Jaccard Index (also known as \gls{iou}) and Manhattan Distance.

The Jaccard Index measures the spatial similarity between prediction and observation. It is defined as the size of the intersection divided by the size of the union of the sample sets, as shown in Equation~\ref{EQN:11}.

\begin{equation}\label{EQN:11}
    \text{Jaccard Index} = \frac{|\A \cap \Ahat|}{|\A \cup \Ahat|}
\end{equation}

where \(\A\) is the simulated or observed reference of the burned area (i.e., ground truth) and \(\Ahat\) denotes the predicted burned area. This metric indicates how well the prediction overlaps with the ground truth.

The Manhattan Distance, also known as the L1 norm, measures the absolute temporal difference in fire spread and is defined as follows:

\begin{equation}\label{EQN:12}
    \text{Manhattan Distance} = \sum_{t=1}^n \left| A_t - \hat{A}_t \right|
\end{equation}

where \(|\A_\t|\) is the number of observed cells on fire at time \(\t\) and \(|\Ahat_\t|\) is number of affected cells in prediction at time \(\t\). This metric indicates how well the prediction matches the ground truth in terms of fire growth and size.

Since the \gls{ca} is stochastic in nature, the same set of parameters can lead to different results. To mitigate this, we obtain target metrics by repeating the simulation of target fires 5 times, and report the average and deviation of the different runs. If, after parameter calibration, our model achieves metrics comparable to the average metrics of the target fire, we consider our parameter calibration a success.

For the real wildfire data, since we can't repeat the fire nor know if the data reflects the time when it was burning. We report only the metrics of our best run. We consider it a success if the fire scale is quantitatively similar to the real fire.

We used a spatial resolution of 30 meters for the side length of a cell, that's the highest resolution from our dataset. Our parameter calibration experiments are conducted using environmental data and fire events located in California, as shown in Table~\ref{TBL:2}. Each fire ecoregion has its unique features. The ecoregion of the 2020 Bear fire is characterized by mountainous terrain, a canyon traversing the map, and an area devoid of vegetation. The ecoregion of the 2020 Brattain fire is expansive, with high altitude and gentle slopes. It features Summer Lake in the northeast, a hill in the west, and overall sparse vegetation density. The 2017 Pier fire ecoregion consists of mountainous terrain with several mountains and hills, relatively steep slopes, and most areas covered with dense vegetation.

\begin{table*}[width=\textwidth,cols=6,pos=h]
    \caption{Overview of selected wildfire datasets, including coordinates, dates, and areas.}
    \begin{tabular*}{\tblwidth}{@{} LLLLLL@{} }
        \toprule
        Dataset name & Latitude & Longitude & Start date & Duration & Area \\
        \midrule
        Bear 2020 & 39.8173 & -121.0893 & 2020-08-19 & 23 days & \(\approx 107 \text{ km}^2\) \\
        Brattain 2020 & 42.6179 & -120.5761 & 2020-09-07 & 22 days & \(\approx 458 \text{ km}^2\) \\
        Pier 2017 & 36.1226 & -118.7074 & 2017-08-29 & 31 days & \(\approx 250 \text{ km}^2\) \\
        \bottomrule
    \end{tabular*}
    \label{TBL:2}
\end{table*}

\begin{figure*}[pos=htbp]
    \centering
    \includegraphics[width=\textwidth]{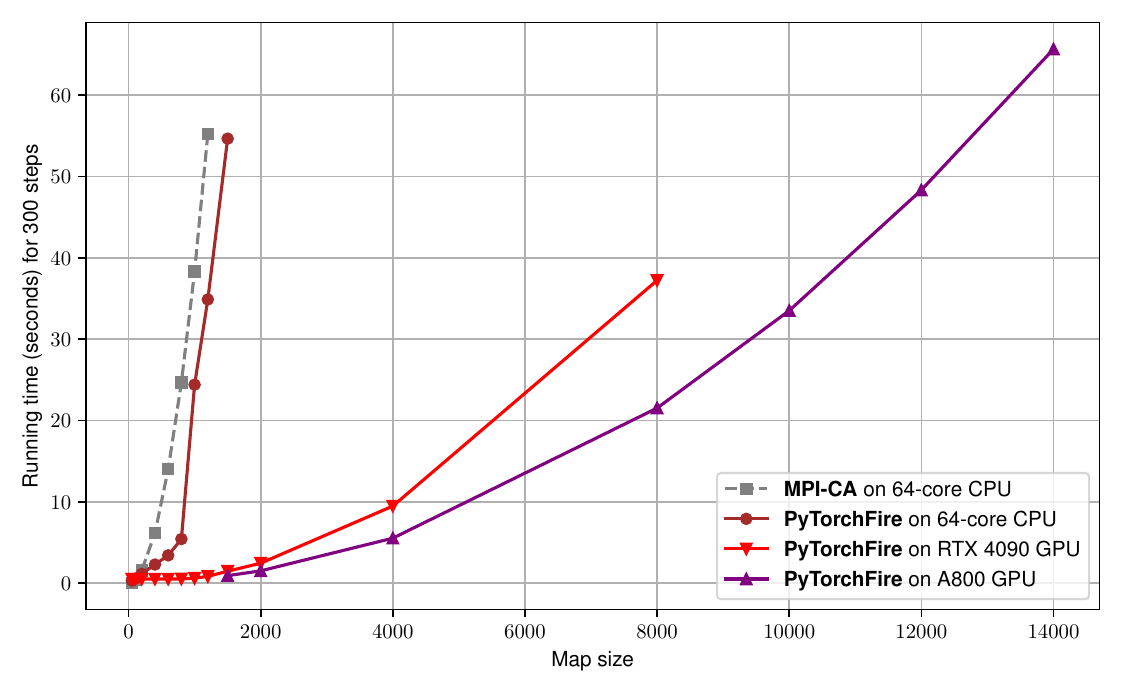}
    \caption{Prediction time (in seconds) for completing 300 steps of simulation across various map sizes (\(50\times14,000\) to \(14,000\times14,000\)). The performance of \texttt{PyTorch} on both CPU and \gls{gpu} outperforms \texttt{MPI-CA}, demonstrating high efficiency and scalability.}
    \label{FIG:6}
\end{figure*}

\textbf{Forward prediction with simulated wildfires.} To evaluate the performance of \texttt{PyTorchFire}, we compared it with another CPU parallel simulator, \texttt{MPI-CA}~\cite{liXCLiParallel_CellularAutomaton_Wildfire2018}, a wildfire \gls{ca} implemented in Python and parallelized using \gls{mpi} techniques. The map in this experiment is a square plain landscape, with fixed wind to mitigate the impact of data loading. We conducted a series of experiments with simulated wildfires running 300 time steps across various map sizes and devices. The results are presented in Figure~\ref{FIG:6}, with the map size representing the side length of the simulation space. Both \texttt{MPI-CA} and \texttt{PyTorchFire} complete the simulation of size 200 in 2 seconds on a CPU\@. However, as the map size increases, there is a rapid escalation in runtime for both simulators. When the map size reaches 1000, which is reasonable after adding buffering areas, both simulators require approximately 30 seconds to complete. We also tested the maximum map size, with \texttt{MPI-CA} handling 1200 and \texttt{PyTorchFire} handling 1500 in one minute.

When utilizing a \gls{gpu}, the performance potential of our solution is fully realized.\ \texttt{PyTorchFire} can complete 300 time steps of simulation for a map size under 1200 in less than 1 second on an NVIDIA\textsuperscript{\tiny\textregistered} GeForce RTX\textsuperscript{\tiny\texttrademark} 4090 \gls{gpu}. With an NVIDIA\textsuperscript{\tiny\textregistered} A800 \gls{gpu}, the simulator can handle 300 time steps for a map size up to 1500 in less than 1 second, proving the simulator's high efficiency and scalability.

For all map sizes under 800 on an NVIDIA\textsuperscript{\tiny\textregistered} GeForce RTX\textsuperscript{\tiny\texttrademark} 4090 \gls{gpu}, we observed a consistent runtime of 0.52 seconds, suggesting that the \gls{gpu} may not be fully utilized for smaller map sizes. We also explored the limits of our simulator. The NVIDIA\textsuperscript{\tiny\textregistered} GeForce RTX\textsuperscript{\tiny\texttrademark} 4090 can support simulations on map sizes up to 8000. With a more advanced \gls{gpu}, \texttt{PyTorchFire} can handle simulations of map sizes up to 14000, completing the entire 300-step simulation in less than a minute on a single \gls{gpu} system.

These results demonstrate that \texttt{PyTorchFire} is highly efficient and scalable, capable of handling large-scale wildfire simulations. This offers significant research potential for developing large-scale wildfire simulations and providing real-time decision support.

\textbf{Inverse modelling with simulated wildfires.} To test the effectiveness of parameter calibration via gradient descent across different scenarios using \texttt{PyTorchFire}, we selected three maps: ``Bear 2020'', ``Brattain 2020'', and ``Pier 2017'' for wildfire simulations. All experiments utilized random \(\pcontinue\) values ranging from [0.4, 0.6], with wind updates at fixed intervals.

To conduct a fundamental test of inverse modeling for both initial parameter underestimates and overestimates, we will demonstrate the performance on two fire ecoregions (maps): \textit{Bear 2020} and \textit{Brattain 2020}. In each map, one simulation time step \(\t\) is equivalent to 1.6 hours in the real world.

To generate a target fire for inversion, initial ignition points of \(3 \times 3\) grid cells and random parameters were manually set. We first ran a fire simulation as the \textit{target} fire, then repeated the same simulation five times to mitigate the randomness of the \gls{ca} simulation. The average and deviation of metrics were recorded as the \textit{metrics of target} fire, which also served as the goal for parameter calibration. Using the initial parameter set (\(\ph = 0.15\) or \(0.8\), with other parameters set to 0), we performed the uncalibrated simulation, recorded as \textit{initial}. Finally, we calibrated the model using the target fire and used the calibrated parameters to perform the simulation, recorded as \textit{calibrated}.

When illustrating the simulated and observed burned area, vegetation and slope were mixed and added to the figure as background, where purple indicates low vegetation or gentle slopes, and green indicates dense vegetation or steep slopes. Time slices were selected in the figure to observe how the simulation changes at different time steps with parameters and how metrics change with calibration. The dynamics of affected cell count was shown to assess the temporal performance of the calibration, and the Jaccard Index was shown to evaluate the stability of the accuracy in different iterations.

\begin{figure*}[pos=htbp]
    \centering
    \includegraphics[width=\textwidth]{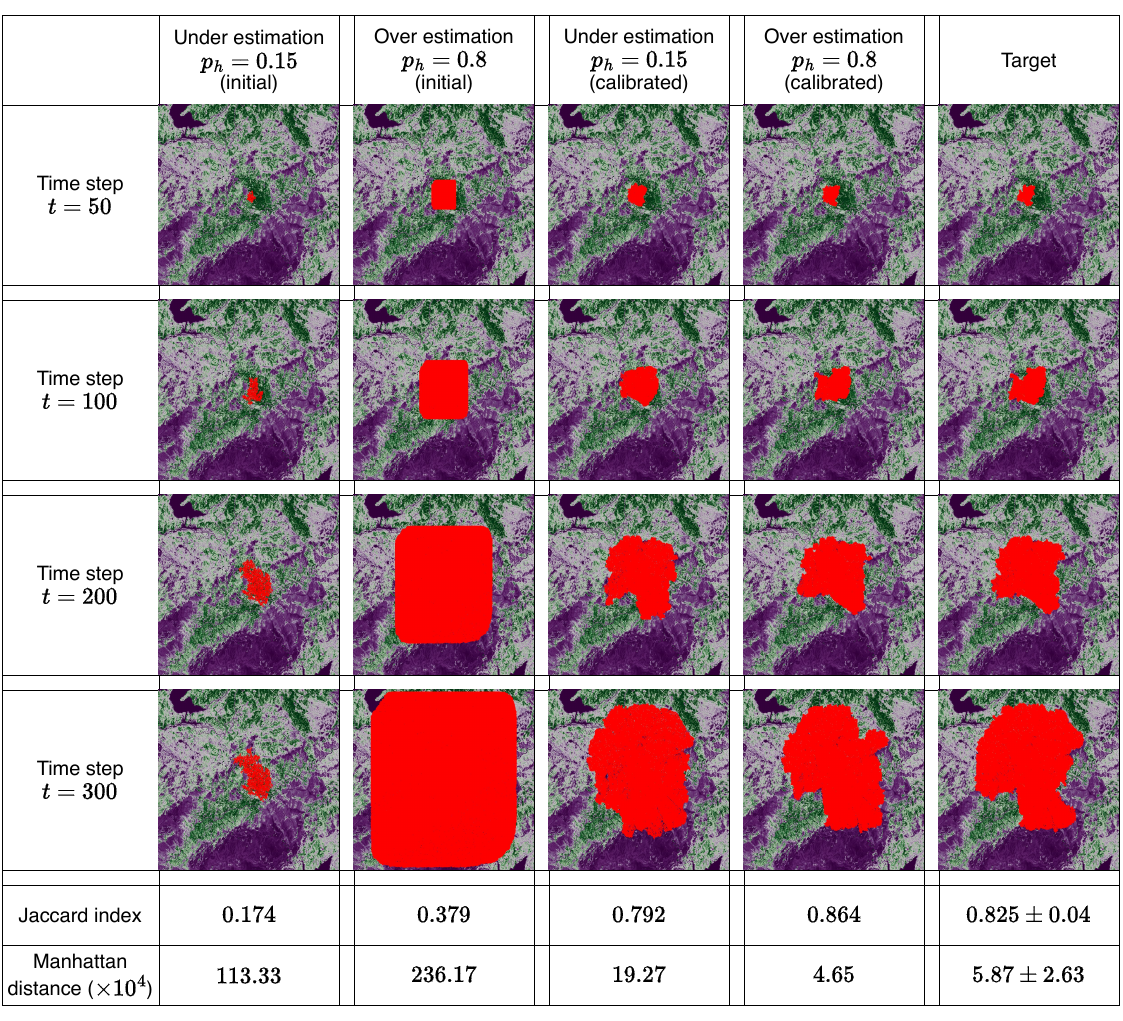}
    \caption{Comparison of wildfire simulation results at various time steps in the Bear 2020 fire ecoregion. Two initial parameter sets (under-estimation and over-estimation) are used. Post-calibration, the simulations closely align with the target fire scale, demonstrating effective recovery despite the inherent randomness of \gls{ca}.}
    \label{FIG:7}
\end{figure*}

\begin{figure*}[pos=htbp]
    \centering
    \includegraphics[width=\textwidth]{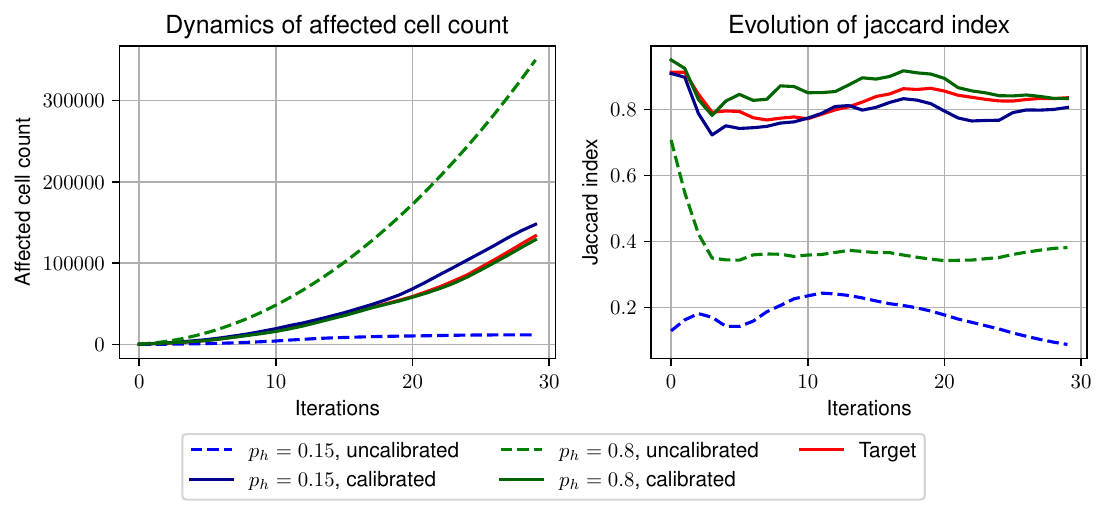}
    \caption{Calibration results in the Bear 2020 fire ecoregion. The `Dynamics of Affected Cell Count' chart shows alignment with targets across various iterations. The `Evolution of Jaccard Index' chart reflects the accuracy stability.}
    \label{FIG:8}
\end{figure*}

As shown in Figure~\ref{FIG:7}, in the ecoregion of ``Bear 2020'', the target grows at a relatively fast speed and gradually forms a shape. However, this shape may not be stable, as the Jaccard Index between the target and other simulations (with same ignition cells) is only 82.5\%. The uncalibrated simulation significantly differs from the target, with Jaccard Indices of 17\% and 37\%, and a Manhattan Distance that is 20 to 40 times larger than the target. After parameter calibration, the simulation aligns with the target fire scale, showing noticeable improvements in both metrics. The metrics for the underestimation case are closer to the target, and the overestimation case even falls within the target range, exhibiting a similar unique shape of fire. From Figure~\ref{FIG:8}, it is evident that the calibrated model's dynamics of affected cell count aligns with the target, and its Jaccard Index remains stable at 0.8 across iterations, both demonstrating significant improvements over the uncalibrated model.

\begin{figure*}[pos=htbp]
    \centering
    \includegraphics[width=\textwidth]{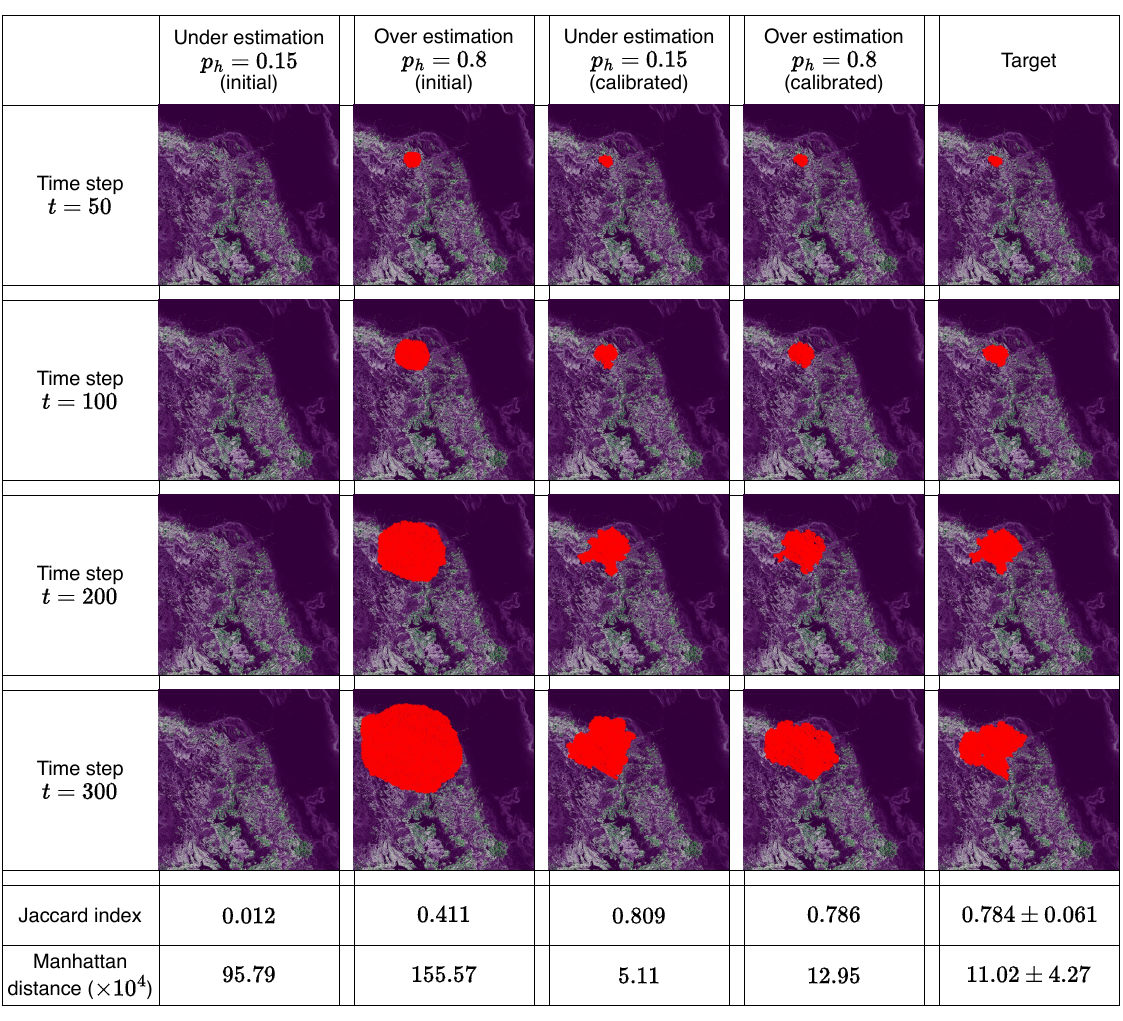}
    \caption{Comparison of wildfire simulation results at various time steps in the Brattain 2020 mfire ecoregion. Two initial parameter sets (underestimation and overestimation) are used. Post-calibration, the simulations closely align with the target fire scale, demonstrating effective recovery despite the inherent randomness of \gls{ca}.}
    \label{FIG:9}
\end{figure*}

\begin{figure*}[pos=htbp]
    \centering
    \includegraphics[width=\textwidth]{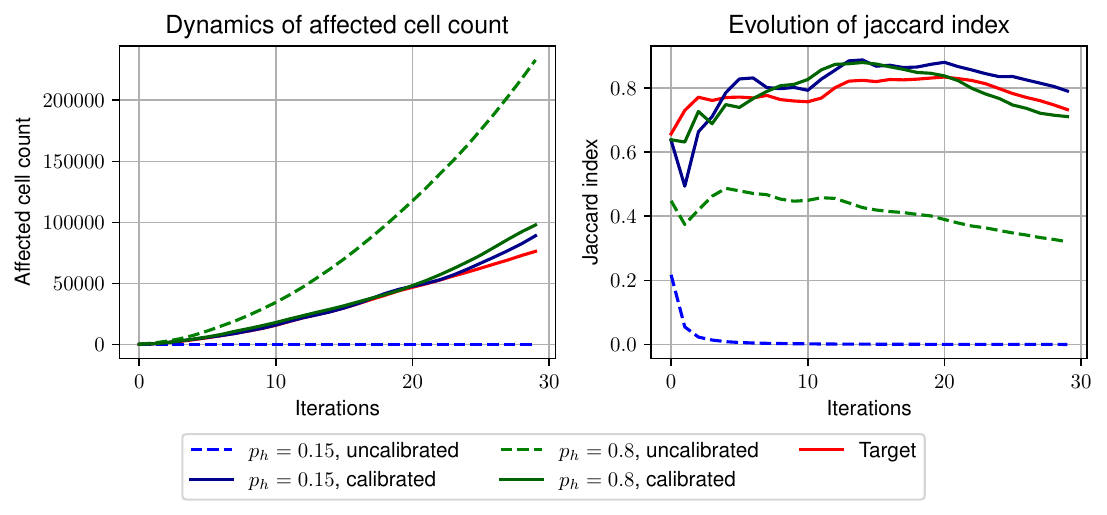}
    \caption{Calibration results in the Brattain 2020 fire ecoregion. The `Dynamics of Affected Cell Count' chart shows alignment with targets across various iterations. The `Evolution of Jaccard Index' chart reflects the accuracy stability.}
    \label{FIG:10}
\end{figure*}

A similar results can be seen in the ``Brattain 2020'' map. As shown in the Figure~\ref{FIG:9}, the target grows slowly, and have a strong spread direction. The target shape is even more unstable with worse metrics (e.g., an average Jaccard Index of \(78.4\%\)) between repeated target runs. The uncalibrated simulation is significantly worse in visualization and metrics, the uncalibrated fire for under estimation case didn't spread at all, with a Jaccard Index of 1\%, and the over estimation case is burning without considering the low density of the vegetation. After calibration, the Jaccard Index has improved from 37\% to 79\%, and the Manhattan Distance has reduced by several orders of magnitude, falls into the range of target. Figure~\ref{FIG:10} shows that the calibrated model has its dynamics of affected cell count aligns with the target, and its Jaccard Index is stable after several initial iterations, both showing great improvement than the uncalibrated one.

To further experiment with our parameter calibration method's effectiveness across different maps and ignition points, we conducted a comparative analysis of wildfire parameter calibration results using multiple simulated datasets. Each dataset was evaluated under two pre-generated fire condition scenarios and five distinct initial ignition points, with the parameters set as: \(a = 0.15\), \(\cone = 0.15\), \(\ctwo = 0.15\). The parameter \(\pcontinue\) was randomly selected within the range \([0.4, 0.6]\), and the parameter \(\ph\) was manually set for each map. Specifically, \(\ph\) values were 0.25 and 0.5 for Bear 2020, 0.5 and 0.6 for Brattain 2020, and 0.45 and 0.6 for Pier 2017.

The comparison of calibrated and uncalibrated fire simulations against the targets (reference simulations) is presented in Table~\ref{TBL:3}. A noticeable improvement in both metrics can be observed across all datasets except for Bear 2020 with a small fire scale. This demonstrates that our method effectively calibrates parameters in most cases. Notably, in the Bear 2020 small-scale fire case, the target itself has a low Jaccard Index of 56\%, compared to approximately 70\% in other cases. This indicates that the small fire on this map is highly random, making it challenging to reliably predict the fire using the current \gls{ca} model. This highlights a limitation of the current fire model equations~\cite{alexandridisCellularAutomataModel2008}, suggesting that an improved fire model is needed to enhance performance in this scenario. Although the running time for parameter calibration varies across different datasets, it remains acceptable for real-time applications, with most cases achieving acceptable parameters after just 3 or even 1 epoch of calibration.

\begin{table*}[width=\textwidth,cols=9,pos=h]
    \caption{Comparative analysis of wildfire parameter calibration results across multiple simulated datasets. Each dataset is evaluated under two pre-generated fire condition scenarios and five different initial ignition points.}
    \begin{tabular*}{\tblwidth}{@{} LLLLLLLLL@{} }
        \toprule
        \multirow{2}{*}{Map name} & \multirow{2}{*}{Scale} & \multicolumn{3}{l}{Jaccard Index} & \multicolumn{3}{l}{Manhattan Distance (\(\times 10^4\))} & Running time (s) \\ \cline{3-5} \cline{6-8} \cline{9-9}
        &  & Uncalibrated & Calibrated & Target & Uncalibrated & Calibrated & Target & Per epoch \\
        \midrule
        \multirow{2}{*}{Bear 2020} & Small & \( 0.441 \pm 0.072 \) & \( 0.452 \pm 0.061 \) & \( 0.563 \pm 0.113 \) & \( 20.34 \pm 12.95 \) & \( 36.56 \pm 20.17 \) & \( 12.73 \pm 8.87 \) & 11.9 \\
        & Large & \( 0.378 \pm 0.144 \) & \( 0.907 \pm 0.065 \) & \( 0.931 \pm 0.018 \) & \( 113.02 \pm 30.32 \) & \( 8.90 \pm 11.64 \) & \( 2.98 \pm 1.94 \) & 26.7 \\
        \multirow{2}{*}{Brattain 2020} & Small & \( 0.021 \pm 0.018 \) & \( 0.733 \pm 0.092 \) & \( 0.767 \pm 0.065 \) & \( 125.62 \pm 28.62 \) & \( 24.67 \pm 12.45 \) & \( 7.32 \pm 4.20 \) & 60.0 \\
        & Large & \( 0.013 \pm 0.005 \) & \( 0.733 \pm 0.039 \) & \( 0.687 \pm 0.162 \) & \( 83.62 \pm 31.09 \) & \( 15.04 \pm 12.83 \) & \( 14.87 \pm 12.95 \) & 63.3 \\
        \multirow{2}{*}{Pier 2017} & Small & \( 0.065 \pm 0.098 \) & \( 0.666 \pm 0.104 \) & \( 0.693 \pm 0.120 \) & \( 40.00 \pm 17.36 \) & \( 12.25 \pm 7.65 \) & \( 9.45 \pm 5.07 \) & 42.4 \\
        & Large & \( 0.071 \pm 0.102 \) & \( 0.765 \pm 0.146 \) & \( 0.774 \pm 0.173 \) & \( 118.94 \pm 27.89 \) & \( 18.47 \pm 10.50 \) & \( 12.27 \pm 14.92 \) & 45.8 \\
        \bottomrule
    \end{tabular*}
    \label{TBL:3}
\end{table*}

\textbf{Experiments with real wildfire data}

To evaluate the effectiveness of parameter calibration in real-world wildfire scenarios, we conducted an experiment analogous to our previous one. Specifically, we measured the relevant metrics both before and after calibration, utilizing real-world fire observation data obtained daily from satellite imagery of MODIS during the fire event~\cite{chengParameterFlexibleWildfire2022}. Additionally, dynamic wind data, updated daily, was incorporated to enhance the simulation environment.

This section presents the visualization of parameter calibration on real-world wildfire datasets `Bear 2020' and `Pier 2017'. In the Bear 2020 map, one simulation time step \( \t \) is equivalent to 0.8 hours in the real world, and in Pier 2017, one simulation time step \( \t \) is equivalent to 1.6 hours in the real world.

\begin{figure*}[pos=htbp]
    \centering
    \includegraphics[width=\textwidth]{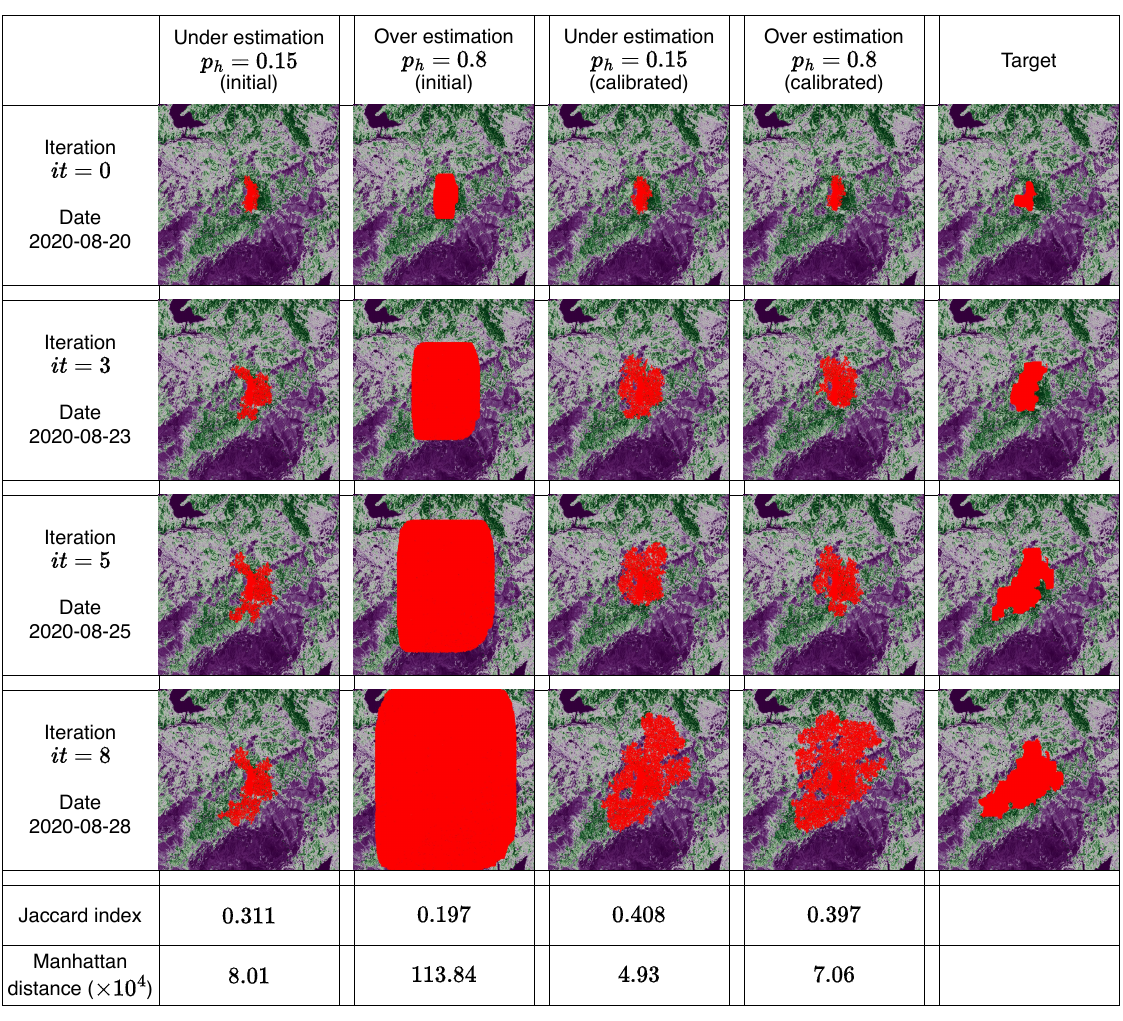}
    \caption{Comparison of wildfire simulation results with real-world satellite observations at various time steps on the Bear 2020 map. Post-calibration, the simulations align with the actual fire scale, demonstrating effective recovery in real-world scenarios.}
    \label{FIG:11}
\end{figure*}

\begin{figure*}[pos=htbp]
    \centering
    \includegraphics[width=\textwidth]{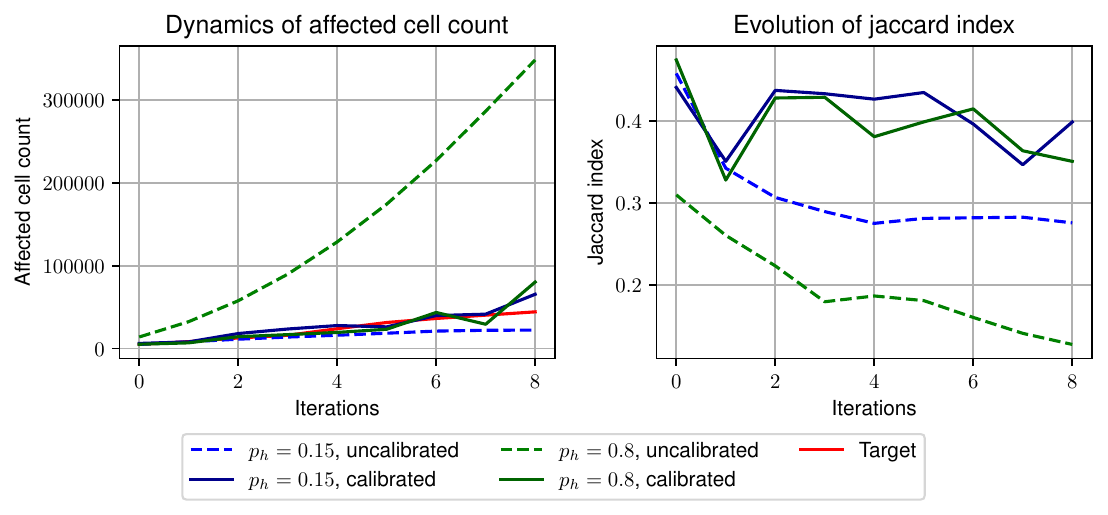}
    \caption{Simulation results after calibration on the Bear 2020 map. The `Dynamics of Affected Cell Count' chart demonstrates alignment with targets across various time steps. The `Evolution of Jaccard Index' chart indicates the accuracy stability of fire scale recovery.}
    \label{FIG:12}
\end{figure*}

Figure~\ref{FIG:11} shows that, in the map `Bear 2020', the target fire grows in a specific direction, gradually forming a shape. However, the growth speed changes from fast to slow over time. The uncalibrated simulation of underestimation case is close to the target in shape but does not match the same growth. The over-estimation fire differs significantly from the target fire. After calibration, the metrics of both cases improve. The underestimation case is closer to the target in fire scale, and the overestimation case now has a similar fire scale to the target fire. It is difficult to achieve the same shape due to the randomness of fire dynamics and the accuracy requirement of the dataset. Figure~\ref{FIG:12} shows that the dynamics of affected cell count align with the target, but it is hard to imitate the gradual growth curve of real-world fire. This is a limitation of the current \gls{ca} fire model, in that the fire tends to spread at a uniform speed during the simulation. The Jaccard Index remains stable and has greatly improved compared to the uncalibrated one.

\begin{figure*}[pos=htbp]
    \centering
    \includegraphics[width=\textwidth]{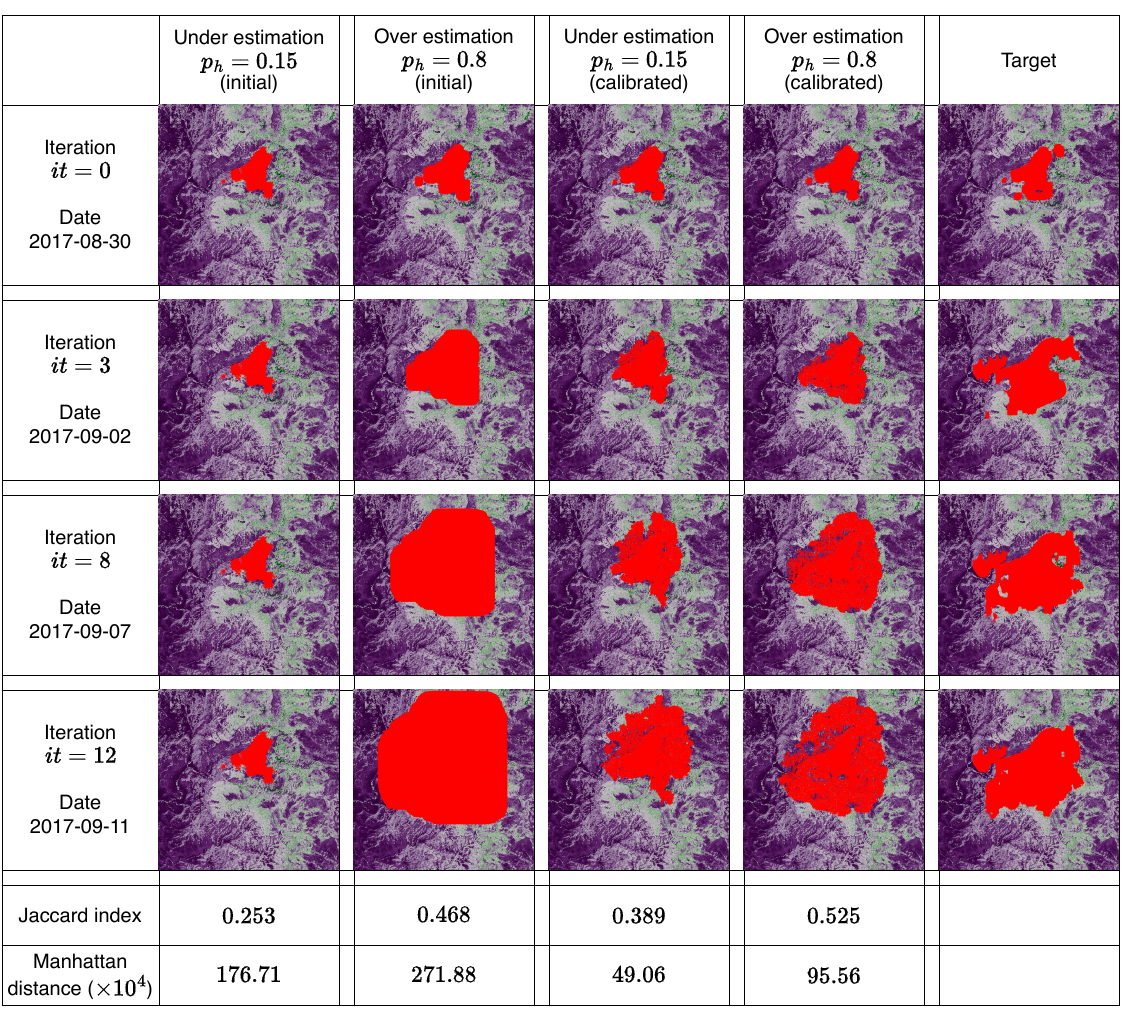}
    \caption{Comparison of wildfire simulation results with real-world satellite observations at various time steps on the Pier 2017 map. Post-calibration, the simulations align with the actual fire scale, demonstrating effective recovery in real-world scenarios.}
    \label{FIG:13}
\end{figure*}

\begin{figure*}[pos=!htb]
    \centering
    \includegraphics[width=\textwidth]{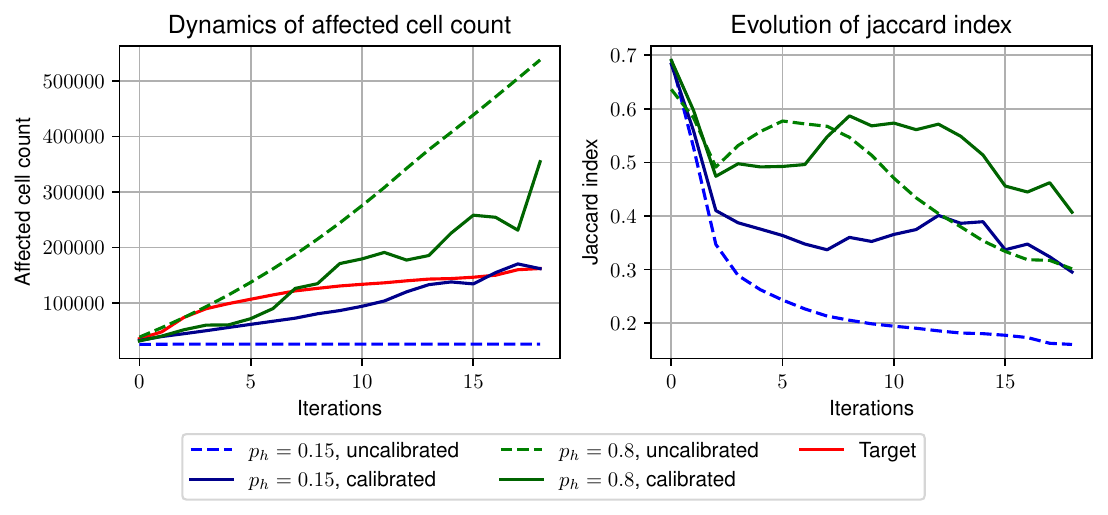}
    \caption{Simulation results after calibration on the Pier 2017 map. The `Dynamics of Affected Cell Count' chart demonstrates alignment with targets across various time steps. The `Evolution of Jaccard Index' chart indicates the accuracy stability of fire scale recovery.}
    \label{FIG:14}
\end{figure*}

Figure~\ref{FIG:13} shows that, in the map `Pier 2017', the target fire grows very slowly after forming a basic shape. The uncalibrated simulation of both cases differs from the target. After calibration, the metrics of both cases improve greatly in visualization. Both cases have a similar shape to the target fire and, most importantly, have their fire scale aligned with the target. The Manhattan Distance has improved by up to three times compared to before. Figure~\ref{FIG:14} shows that the dynamics of the affected cell count become closer to the target. However, it is challenging to perfectly imitate the gradual growth curve of real-world fire due to the limitations of the \gls{ca} model. The Jaccard Index remains stable and has greatly improved compared to the uncalibrated one.

\section{Conclusions}\label{SEC:CONCLUSION}

In conclusion, \texttt{PyTorchFire} represents a significant advancement in wildfire simulation and parameter calibration, leveraging the computational power of \glspl{gpu} and the flexibility of \texttt{PyTorch}.\ \texttt{PyTorchFire} utilizes geophysical and climate data as input to predict the spatio-temporal spread of wildfires. The software is fully scalable to accommodate different resolutions of input and observation data. By introducing the redesigned differentiable wildfire \gls{ca}, \texttt{PyTorchFire} achieves remarkable computational efficiency, enabling millisecond-level simulations and real-world-scale wildfire modeling at high resolution, which are essential for timely wildfire management. The integration of automatic differentiation and gradient descent for parameter calibration ensures that the simulations closely align with real-world wildfire behavior, enhancing both the accuracy and reliability of predictions.\ \texttt{PyTorchFire}'s ability to maintain the stochastic nature of fire propagation while operating in real-world ecoregions makes it an invaluable tool for researchers and practitioners in the field of wildfire management. Its generalized and environment-data-sensitive design performs well in simulating early-stage fires, contributing to future wildfire early warning systems. As an open-access software, \texttt{PyTorchFire} not only advances the state-of-the-art in wildfire simulation but also democratizes access to powerful computational tools, fostering innovation and collaboration in combating uncontrolled wildfires. It has the potential to be integrated into automated fire monitoring system toolchains.

While current wildfire warnings primarily focus on the probability of ignition\cite{scottWildfireRiskAssessment2013}, they often overlook the spatial dynamics and potential consequences of the fire's spread. This probabilistic approach tends to prioritize areas based on their likelihood of ignition rather than the potential severity or impact of a fire once it occurs. Incorporating spatial extent into risk assessments could provide a more holistic view of wildfire threats. By considering factors such as fuel, topography, and weather patterns, we can better predict not only where a fire might start but also how it might evolve and spread across the landscape. This expanded perspective could enhance operational strategies, allowing for more targeted resource allocation and improved mitigation efforts. Such an approach acknowledges the complex interplay between environmental variables and human factors, ultimately leading to a more resilient response framework. By shifting our focus from mere probabilities to comprehensive risk evaluations, we can better safeguard communities and ecosystems from the multifaceted threats posed by wildfires.

Future work should focus on extending the application of \texttt{PyTorchFire} to global wildfire scenarios, incorporating various human interventions such as fire retardant placement and controlled burns. In this paper, we present a proof of concept for \acrfull{dca} with \gls{gpu} acceleration, rather than an operational model. Our proposed framework is flexible and can be easily adapted for more advanced \gls{ca} models, such as PROPAGATOR~\cite{trucchiaPROPAGATOROperationalCellularAutomata2020}. For operational improvements, integrating these models and incorporating additional parameters, such as fine fuel moisture content, can enhance the realism of simulations. These parameters can be added by applying and calibrating appropriate scaling factors to Equation~\ref{EQN:2}. Reliable conversion of landscape data to scaling factors will significantly enhance the simulation. Additionally, due to the limitations of \gls{ca}, methods need to be developed to achieve a non-linear spread speed. Obtaining high-resolution wind and fire data can also improve the accuracy of the simulation and parameter calibration. Once spotting effects are implemented, the simulation of fire dynamics will be more accurate. Modeling multiple ignition points is also pending.

\appendix\label{SEC:APPENDIX}

\section{Dataset}\label{APP:DATASET}

All data in this paper were projected and then cropped using the EPSG:3310 coordinate system for a consistent spatial reference. To avoid erroneous edge effects, a 4980-meter padded boundary was added to the original fire-affected region.

For landscape data, we utilized resources from \texttt{LANDFIRE}, downloaded via the LANDFIRE Product Service. Specifically, we used the Forest Canopy Bulk Density Version 2.3.0~\cite{LANDFIRE230} (unit: \(\text{kg/m}^3 \times 100\)) as our density input, the Forest Canopy Cover Version 2.3.0~\cite{LANDFIRE230} (unit: percentage) for canopy data, and the Slope 2020 dataset~\cite{LANDFIRE220} (in degrees) for slope data.

Wind data were obtained from the Google Earth Engine dataset \texttt{ECMWF/ERA5\_LAND/DAILY\_AGGR}~\cite{joaquinERA5LandMonthlyAveraged2019}. The \linebreak \texttt{u\_component\_of\_wind\_10m} and \texttt{v\_component\_of\_wind\_10m} bands were used to compute wind velocity and direction.

We used fire satellite observation data generated from \gls{modis}~\cite{giglioCollectionMODISActive2016}.\ \gls{modis} provides thermal observations globally four times a day (Terra at 10:30 and 22:30; Aqua at 13:30 and 01:30 local time) at a resolution of approximately 1 km.

\section{Environment}\label{APP:ENVIRONMENT}

Most experiments were conducted on a shared server equipped with 64 vCPUs from an AMD EPYC\textsuperscript{\tiny\texttrademark} 9654 @ 2.40GHz (with AVX512 support), 240GB of memory, and four NVIDIA\textsuperscript{\tiny\textregistered} GeForce RTX\textsuperscript{\tiny\texttrademark} 4090 (24GB) graphics cards. The operating system used was Ubuntu 22.04.3 LTS x86\_64. The software environment included \texttt{PyTorch} 2.3.1 running on Python 3.11 (Ubuntu 22.04) with \gls{cuda} version 12.4.

To further evaluate the program's potential, additional tests were performed on a server with 14 vCPUs from an Intel\textsuperscript{\tiny\textregistered} Xeon\textsuperscript{\tiny\textregistered} Gold 6348 CPU @ 2.60GHz, 100GB of memory, and one NVIDIA NVIDIA\textsuperscript{\tiny\textregistered} A800 80GB PCIe graphics card. The operating system for this setup was also Ubuntu 22.04.3 LTS x86\_64. The software environment remained consistent, with \texttt{PyTorch} 2.3.1 on Python 3.11 (Ubuntu 22.04) and \gls{cuda} version 12.4.

\section{Learning rate selection}\label{APP:LR}

To obtain optimal results, we experimented with multiple learning rates. Our findings indicate that the optimal learning rate varies depending on the specific map. However, they do not vary significantly in terms of results after sufficient epochs of calibration. We recommend that users experiment with different learning rates to determine the one that yields the best outcome.

For cases in Figure~\ref{FIG:7}, the learning rate is \(5 \times 10^{-3}\) for the under-estimation case and \(1 \times 10^{-2}\) for the over-estimation case. For cases in Figure~\ref{FIG:9}, the learning rate is \(5 \times 10^{-3}\) for the under-estimation case and \(7 \times 10^{-3}\) for the over-estimation case.

In Figure~\ref{FIG:11}, the learning rate is \(1 \times 10^{-2}\) for both under-estimation and over-estimation cases. In Figure~\ref{FIG:13}, the learning rate is \(5 \times 10^{-3}\).

For the rest of the experiments, the learning rate is \(5 \times 10^{-3}\).

\printglossary[type=\acronymtype, title=Abbreviations]

\printcredits

\section*{Declaration of competing interest}
The authors declare that they have no known competing financial interests or personal relationships that could have appeared to influence the work reported in this paper.

\section*{Declaration of generative AI and AI-assisted technologies in the writing process}
During the preparation of this work, the authors utilized OpenAI's \texttt{gpt-4o-2024-05-13} API to enhance readability and language. Following the use of this service, the authors reviewed and edited the content as necessary and assume full responsibility for the content of the publication.

\section*{Acknowledgments}
The authors would like to thank Juntao Lu, Yifan Shao, and Tianle Zhong (University of Virginia) for their valuable discussions on the research direction. We also extend our gratitude to Gang Xia for providing feedback on this paper. This paper is dedicated to the memory of the first author's grandfather, Junhong Xia. The authors acknowledge the support of the French Agence Nationale de la Recherche (ANR) under reference ANR-22-CPJ2-0143-01. This research was supported by the academic research grant from NVIDIA\@.

\balance

\bibliographystyle{elsarticle-num}

\bibliography{main.bib}

\end{document}